# A Day-ahead Market Energy Auction for Distribution System Operation

M. Nazif Faqiry, A. Khaled Zarabie, Fatehullah Nassery, Hongyu Wu, Sanjoy Das

# A Day-ahead Market Energy Auction for Distribution System Operation

M.N. Faqiry, A.K.Zarabie, F.Nassery, Hongyu Wu, Sanjoy Das
Electrical & Computer Engineering, Kansas State University
Corresponding Author (mnfaqiry@ksu.edu)

*Abstract*—In this paper, we study a Day-ahead Market(DAM) double energy auction in a distribution system involving dispatchable generation units, renewable generation units supported by battery storage systems(BSSs), fixed loads, price responsive loads, and supply from the Wholesale Market(WSM) at Locational Marginal Price (LMP). The auction is implemented within a Distribution System Operator (DSO) premises using Mixed Integer Linear Programming (MILP). The proposed auction is cleared at the Distribution LMP (DLMP) and is observed to be weakly budget balanced if no penalty is applied for DSO's deviation from its original commitment with the WSM. Furthermore, the dynamics of DLMP versus LMP, and their effect on distribution market participants scheduled quantities as well as the WSM supply to the distribution system is investigated.

## NOMENCLATURE

Indices:

$m$   Distribution bus index
$g$   Generation unit index
$b$   BSS unit index
$l$   Load index
$q$   Generation segments index
$r$   Load segments index
$t$   Timeslot index

Sets:

$\mathcal{M}$   Set of distribution buses
$\mathcal{G}$   Set of generation units
$\mathcal{B}$   Set of BSS units
$\mathcal{L}$   Set of price responsive loads
$\mathcal{T}$   Set of time slots

Functions:

$f$   Objective function indicating social welfare
$\phi$   Deviation from commitment penalty function

Variables:

$px$   Segment generation
$p$    Generation unit output
$e$    BSS unit energy output
$s$    Supply from WSM
$dx$   Segment load
$d$    Total demand of load
$c$    BSS unit state of charge

Binary and Integer Variables:

$i$    Commitment state of generation unit
$j$    Commitment state of BSS unit
$y$    Startup indicator of generation unit
$z$    Shut down indicator of generation unit
$sd$   Shut down counter for generation unit
$su$   Start up counter for generation unit
$u$    BSS unit charging indicator
$v$    BSS unit discharging indicator
$cc$   Charging counter for BSS unit
$dc$   Discharging counter for BSS unit

Parameters:

$S$          Total fixed supply allocated by the ISO to DSO
$PX^{max}$   Maximum segment generation in each segment
$P^{min}$    Minimum generation of a generation unit
$P^{max}$    Maximum generation of a generation unit
$CB$         Selling cost of BSS unit energy
$CG$         Selling cost of generation unit energy
$CL$         Buying cost of load
$STC$        Start-up cost of a generation unit
$SDC$        Shut down cost of a generation unit
$RU$         Rump up rate of a generation unit
$RD$         Rump down rate a generation unit
$MDTG$       Minimum down time of a generation unit
$MUTG$       Minimum up time of a generation unit
$MDTB$       Minimum discharge time of a BSS unit
$MCTB$       Minimum charge time of a BSS unit
$E^{min}$    Minimum BSS unit energy withdraw amount
$E^{max}$    Maximum BSS unit energy withdraw amount
$C^{min}$    Minimum state of charge of BSS unit
$C^{max}$    Maximum state of charge of BSS unit

## I. INTRODUCTION

The study of market design for distribution system operator (DSO) has recently gained a considerable research momentum due to inclusion of different market participants such as Renewable Energy Resource (RES) owners, Battery Storage System (BSS) owners, and load aggregators in the smart grid. As an intermediate market operator, the DSO may use forecasted and (or) historical load and system distributed generation (DG) data to bid in the WSM. The independent system operator (ISO) clears WSM and announces the DSO's LMP and scheduled amount. In such a situation, the DSO may implement a DAM auction in its own service territory to seek further efficient resource allocation and maximize social welfare. The service area under the control of the DSO can be comprised of various loads and generation units. The generation units in the network can be of dispatchable and non-dispatchable kind. Non-dispatchable units that are mainly renewable energy resource such as wind and solar are intermittent and causes uncertainty while weakening the classical demand price correlation [1], [2]. However, these restrictions can be alleviated by channeling their output power through BSSs [3], [4].

While distinctive models for DSO are proposed by various researchers in the electricity market [5] − [8], a broader model that can handle involvement of all kinds of market participants has not yet been developed. Distribution market clearing and payment mechanisms are still open questions that are yet to be answered with viable and sound assumptions.

Seen hierarchically, the distribution system service territory starts at the bus where the utility company can bid in the WSM through a DSO. In many cases, a sub-transmission network then distributes the power to different distribution buses (D-buses), i.e. substations [5]. Each D-bus serves smaller substations at medium voltage that may cover a small geographic area or a community of houses at low voltage. In the low voltage distribution system, different aggregator models have been proposed in the literature that may be incentivized to aggregate classical fixed as well as price responsive loads and bid in the distribution market [9] − [12].

This paper proposes a DAM model for DSO given the LMP and its commitment post-WSM or based on forecasted or historical data. The goal of the DSO is to implement a double auction in order to maximize the social welfare of the market participants such as distribution level generation units, BSS units, and price responsive loads through aggregators.

## II. MODEL

In the proposed model when the DSO receives its committed energy information at LMP from the ISO, it asks for bids from generation units, BSSs, and load aggregators in the distribution system. Generation units are assumed to submit a three-segment bid and amount as well as their ramp up/down rates and startup/shut down costs. Load aggregators are assumed capable of dividing their aggregated load to a fixed load segment amount that needs to be served at all times and a two-segment price responsive bid and amount. The fixed segment of the loads are served at the market-clearing price and do not accompany any monetary bid amount. This is because not all loads are price responsive, i.e. a high portion of the load is price inelastic and needs to be served at all times. It is assumed that renewable energy resources at the distribution level are coupled with BSS units for a smooth participation in the auction, and declare their selling price as well as their unit characteristics to the DSO for optimal operation. BSS units have the potential to help integrate deeper penetrations of renewable energy into electricity grids and deliver efficient, low-cost, fundamental electricity-grid services [3]. It is considered in this paper that the BSS units are backed up and charged only by its own renewable energy resource(s). The proposed model can be easily extended for idle BSS units that can also be charged or discharged by the grid.

Equipped with the aforementioned considerations, the DSO runs the auction and clears the market by providing power to the successful biding parties at the DLMP. The concept of DLMP has been proposed by many researchers for distribution system congestion management, market clearing, and loss minimization [2], [6], [13] − [15], [18], as it shows the true marginal cost of supplying the next increment of load.

Although in this study the distribution line congestion has not been taken into account, the true value of a unit of energy at the D-buses are different from that of the DSO bus. This is because bids of the market participants at each D-bus is different, resulting to different DLMPs. The DLMPs are attained by the Lagrange multiplier of the supply-demand balance constraint of each D-bus. The current model can be easily extended to include line congestion constraints using shift factors or Power Transfer Distribution Factor (PTDF) using DC power flow, a viable approximation at the sub-transmission level of the distribution system. Furthermore, one can also use the simplified DistFlow equations [19], [20] to model line congestion as well as D-bus voltage limit constraints.

Fig. 1 depicts a sample distribution system architecture with four D-buses fed by sub-transmission lines from the main DSO bus in a radial configuration. Each D-bus is considered to include dispatchable generation units (denoted by G), BSS units charged by renewable resources (denoted by B) and several aggregated loads (denoted by L). After WSM clearing of DAM, or by means of forecasting or historical data, the DSO knows the committed supply $S_t$ and LMP of the WSM and runs a day−ahead auction to determine its own unit commitment and supply distribution while maximizing the overall system social welfare.

The social welfare maximization of the system can be modeled as in Eqn. (1) subject to system constraints in Eqns. (2) − (30).

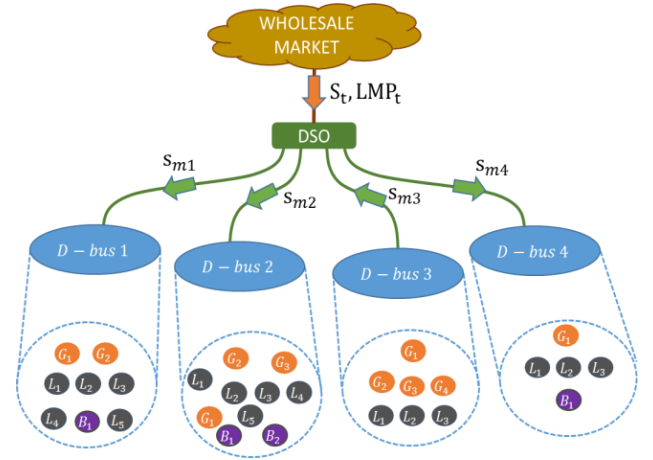

Fig. 1. System architecture with four distribution buses

Maximize $f$:

$$f = \sum_t \sum_m \sum_l \sum_{r>1} \text{CL}_{mlrt}\, \text{dx}_{mlrt} - \sum_t \sum_m \sum_g \sum_q \text{CG}_{mgq}\, \text{px}_{mgqt}$$
$$- \sum_t \sum_m \sum_g \text{STC}_{mg} y_{mgt} - \sum_t \sum_m \sum_g \text{SDC}_{mg} z_{mgt}$$

$$-\sum_t \sum_m \sum_b CB_{mb}\, e_{mbt} - \sum_t \sum_m LMP_t\, s_{mt} \quad (1)$$

subject to the following constraints $\forall m \epsilon \mathcal{M}, \forall g \epsilon \mathcal{G}, \forall b \epsilon \mathcal{B}, \forall l \epsilon \mathcal{L}, \forall t \epsilon \mathcal{T}$:

$$\sum_m s_{mt} \leq S_t \qquad \forall t \quad (2)$$

$$\sum_g p_{mgt} + \sum_b e_{mbt} + s_{mt} = \sum_l d_{mlt} \qquad \forall m, \forall t \quad (3)$$

The first term in the RHS of Eqn. (1) pertains to the price responsive loads and allocates to loads with highest bids. Note that, as the loads are assumed to submit their fixed load segment without and two price responsive segments with monetary bids, the fixed segment is excluded from the first term, i.e. $r > 1$. The second, third, and fourth terms relate to generation units and allocates to those that have bid the least and incur smaller start up and shut down costs. The fifth term is modeled to pick BSS units with lowest bids and the sixth term decides the optimum amount of supply to each D-bus. The objective function in Eqn. (1) picks energy sellers with least marginal cost and energy buyers with highest valuation in order to maximize surplus and result an efficient allocation [10].

Eqn. (2) indicates that the sum of supplies channeled through the DSO into each D-bus shall not exceed the committed schedule to the ISO. Similarly, Eqn. (3) ensures that the demand of each individual D-bus is supplied by the allocated portion of supply from WSM, and the supply of the generation and BSS units in that bus.

$$d_{mlt} = \sum_r dx_{mlrt} \qquad \forall l, \forall m, \forall t \quad (4)$$

$$0 \leq dx_{mlrt} \leq DX_{mlr}^{max} \qquad \forall r, \forall l, \forall m, \forall t \quad (5)$$

$$D_{ml}^{min} \leq d_{mlt} \leq D_{ml}^{max} \qquad \forall l, \forall m, \forall t \quad (6)$$

$$p_{mgt} = \sum_q p_{mgqt} \qquad \forall g, \forall m, \forall t \quad (7)$$

$$0 \leq p_{mgqt} \leq PX_{mgq}^{max} \qquad \forall q, \forall g, \forall m, \forall t \quad (8)$$

$$P_{mg}^{min}\, i_{mgt} \leq p_{mgt} \leq P_{mg}^{max}\, i_{mgt} \qquad \forall g, \forall m, \forall t \quad (9)$$

Eqn. (4) shows that the total demand of a load is equal to the demand of the fixed segment ($r = 1$) plus the demand of its responsive segments ($r > 1$). Eqn. (5) and (6) indicates the segment limits and total demand limits of each load. Eqn. (7), signifies that power generated by a generation unit is equal to the aggregated segment generation of that unit. Eqn. (8) assures that the power generated at each segment by a generation unit does not violate the predefined lower and upper limits of generation in that segment. If committed, the total power generated by a dispatchable unit shall lie within its lower and upper generation limits, as indicated by Eqn. (9).

$$p_{mgt} - p_{mg(t-1)} \leq RU_{mg} \qquad \forall g, \forall m, \forall t \quad (10)$$

$$p_{mg(t-1)} - p_{mgt} \leq RD_{mg} \qquad \forall g, \forall m, \forall t \quad (11)$$

$$0 \leq su_{mgt} \leq |\mathcal{T}|\, i_{mgt} \qquad \forall g, \forall m, \forall t \quad (12)$$

$$(|\mathcal{T}| + 1)i_{mgt} - |\mathcal{T}| \leq su_{mgt} - su_{mg,t-1} \leq 1 \quad \forall g, \forall m, \forall t \quad (13)$$

$$su_{mgt} \geq MUTG_{mg}\, z_{mg,t+1} \qquad \forall g, \forall m, \forall t \quad (14)$$

$$0 \leq sd_{mgt} \leq |\mathcal{T}|\, (1 - i_{mgt}) \qquad \forall g, \forall m, \forall t \quad (15)$$

$$1 - (|\mathcal{T}| + 1)i_{mgt} \leq sd_{mgt} - sd_{mg,t-1} \leq 1 \qquad \forall g, \forall m, \forall t \quad (16)$$

$$sd_{mgt} \geq MDTG_{mg}\, y_{mg,t+1} \qquad \forall g, \forall m, \forall t \quad (17)$$

$$i_{mgt} - i_{mgt-1} = y_{mgt} - z_{mgt} \qquad \forall g, \forall m, \forall t \quad (18)$$

$$y_{mgt} + z_{mgt} \leq 1 \qquad \forall g, \forall m, \forall t \quad (19)$$

The ramp up/down constraints of each individual generation unit is ensured by Eqns. (10) and (11). In the same way, Eqns. (12) – (19) shows the MILP formulation for minimum uptime and minimum down time constraints of the generation units.

$$E^{min} j_{mbt} \leq e_{mbt} \leq E^{max} j_{mbt} \qquad \forall b, \forall m, \forall t \quad (20)$$

$$C^{min} \leq c_{mbt} \leq C^{max} \qquad \forall b, \forall m, \forall t \quad (21)$$

$$c_{mbt} = c_{mb(t-1)} - e_{mbt} \qquad \forall b, \forall m, \forall t \quad (22)$$

$$0 \leq dc_{mbt} \leq |\mathcal{T}|\, j_{mbt} \qquad \forall b, \forall m, \forall t \quad (23)$$

$$(|\mathcal{T}| + 1)j_{mbt} - |\mathcal{T}| \leq dc_{mbt} - dc_{mb,t-1} \leq 1 \qquad \forall b, \forall m, \forall t \quad (24)$$

$$dc_{mbt} \geq MDTB_{mb}\, u_{mb,t+1} \qquad \forall b, \forall m, \forall t \quad (25)$$

$$0 \leq cc_{mbt} \leq |\mathcal{T}|\, (1 - j_{mbt}) \qquad \forall b, \forall m, \forall t \quad (26)$$

$$1 - (|\mathcal{T}| + 1)j_{mbt} \leq cc_{mbt} - cc_{mb,t-1} \leq 1 \qquad \forall b, \forall m, \forall t \quad (27)$$

$$cc_{mbt} \geq MDTB_{mb}\, v_{mb,t+1} \qquad \forall b, \forall m, \forall t \quad (28)$$

$$j_{mbt} - j_{mbt-1} = v_{mbt} - u_{mbt} \qquad \forall b, \forall m, \forall t \quad (29)$$

$$u_{mbt} + v_{mbt} \leq 1 \qquad \forall b, \forall m, \forall t \quad (30)$$

In order to increase the life expectancy of the BSS units, minimum and maximum amount of energy withdrawal from these units are bounded by the given limits in Eqn. (20). Likewise, Eqn. (21) keeps the BSS safe from overcharging and deep discharging. Eqn. (22) indicates the BSS units' charging state update. Consecutive minimum discharging and charging hours' constraints of BSS units are represented by Eqns. (23) - (30).

III. SIMULATION RESULTS

The tabulated data in this section pertains to market participants bidding information and serves as basis for the simulation reported. Table I. shows each BSS units' bidding information submitted to the DSO. The bidding information for each generation unit is summarized in Table II and the hourly WSM supply and LMP is shown in Table III. Loads' bidding information are not included due to space limitations as load bids were assumed to vary over timeslots.

The MILP model presented in Eqns. (1) - (30) is coded in GAMS and solved using CPLEX solver for 24-hour horizon. The simulation results reported corroborate the validity of the proposed model. Several sets of analysis were conducted to see the effect of LMP on the auction outcome. Simulation results indicates that loads, generation and BSS units are very responsive to changes in LMP at the DSO bus at a given amount of supply by the WSM. At low LMPs, more from the WSM is allocated to loads due to high local generation bids. As the LMP increases, more internal generation at each D-bus is scheduled. A similar effect is observed on serving the price responsive loads.

TABLE I. Bidding information of each BSS unit at each D-bus

| Bus | Unit | $(C^{min}, C^{max})$ (MWh) | $(E^{min}, E^{max}]$ (MWh/hr) | (MDTB, MCTB) (hr) | CB ($/MWh) |
|---|---|---|---|---|---|
| 1 | BSS1 | $(1-10)$ | $(0.4-2)$ | $(3-6)$ | 35 |
| 2 | BSS1 | $(1-8)$ | $(0.4-2)$ | $(3-6)$ | 33 |
| 2 | BSS2 | $(1-10)$ | $(0.4-2)$ | $(3-6)$ | 36.5 |
| 4 | BSS1 | $(1-8)$ | $(0.4-2)$ | $(2-6)$ | 34 |

TABLE II. Segment generation and unit price for each D-bus

| (Bus, Unit) | $(PX_1^{max}, CX_1)$ (MW,$) | $(PX_2^{max}, CX_2)$ (MW,$) | $(PX_3^{max}, CX_3)$ (MW,$) | (RU, RD) (MW/h) | (STC, SDC) ($) |
|---|---|---|---|---|---|
| (1,G1) | (1.5, 36.7) | (2.5, 39.3) | (1, 42) | (2.5, 2.5) | (75, 60) |
| (1,G2) | (1.6, 34.8) | (2, 37.8) | (1.4, 40.5) | (2.5, 2.5) | (60, 60) |
| (2,G1) | (1.5, 30) | (1.7, 33) | (1.8, 39) | (2.5, 2.5) | (45, 54) |
| (2,G2) | (1.4, 36.9) | (1.8, 39.6) | (1.8, 43.8) | (2.5, 2.5) | (51, 45) |
| (2,G3) | (1, 34.5) | (1.5, 36) | (0.5, 39.6) | (3, 3) | (84, 45) |
| (3,G1) | (1.2, 29.4) | (1.8, 30.6) | (2, 34.5) | (2.5, 2.5) | (0, 0) |
| (3,G2) | (1.8, 32.1) | (1.45, 32.6) | (1.75, 34.5) | (2.5, 2.5) | (45, 51) |
| (3,G3) | (0.8, 35.7) | (1.7, 37.5) | (0.5, 40.5) | (3, 3) | (60, 48) |
| (3,G4) | (0.95, 36.3) | (1.1, 37.5) | (0.95, 40.5) | (3, 3) | (0, 0) |
| (4,G1) | (1.9, 37.5) | (1.7, 41.4) | (1.4, 44.5) | (2.5, 2.5) | (10, 10) |

TABLE III. Supply at the DSO bus from the ISO at the LMP

| LMP and ISO supply in 24hr scheduling horizon | | | | | | |
|---|---|---|---|---|---|---|
| t | 1 | 2 | 3 | 4 | 5 | 6 |
| LMP | 22.07 | 24.83 | 24.83 | 23.45 | 24.83 | 24.83 |
| $S_t$ | 29.04 | 32.67 | 32.67 | 30.855 | 32.67 | 32.67 |
| t | 7 | 8 | 9 | 10 | 11 | 12 |
| LMP | 26.21 | 27.59 | 28.97 | 30.35 | 33.11 | 30.35 |
| $S_t$ | 34.485 | 36.3 | 38.115 | 39.93 | 43.56 | 39.93 |
| t | 13 | 14 | 15 | 16 | 17 | 18 |
| LMP | 27.59 | 26.21 | 26.48 | 25.38 | 27.04 | 30.35 |
| $S_t$ | 36.3 | 34.485 | 34.848 | 33.396 | 35.574 | 39.93 |
| t | 19 | 20 | 21 | 22 | 23 | 24 |
| LMP | 31.73 | 34.49 | 31.73 | 28.97 | 26.21 | 23.45 |
| $S_t$ | 41.475 | 45.375 | 41.745 | 38.115 | 34.485 | 30.855 |

Fig. 2 depicts LMP versus DSO's supply allocation to each D-bus as a portion of the total committed schedule that it receives from WSM during 24h scheduling horizon. Notice that supply to each D-bus is very responsive to the changes in LMP. When LMP is low, more power is allocated to flow from the WSM. However, during high LMP hours the power flow from DSO drops significantly, to the extent that D-bus number three feeds power back into the distribution network, i.e. to other D-buses. This is because D-bus 3 has more cheap generation and loads in other D-buses are ready to purchase at higher price than that of its own local loads.

Fig. 3 shows LMP versus each D-bus's internal generation. The internal generation at each D-bus increases with increase in LMP. Fig. 4 shows total power demand equals the total supply, which consists of supply from the WSM, internal generation, and supply from BSS units. Notice that the BSS units are scheduled only during peak LMP hours for at least three consecutive hours due to their minimum discharging time constraints.

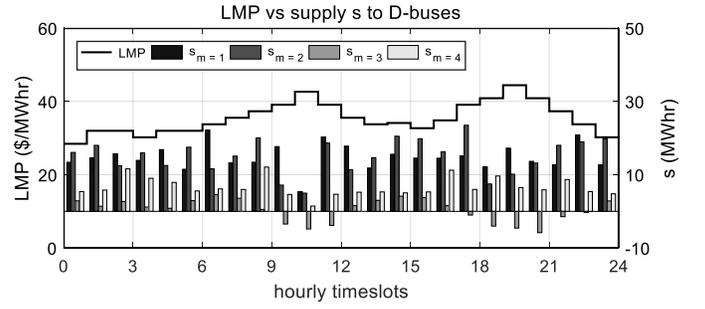

Fig. 2. Hourly DSO supply to each D-bus vs. LMP

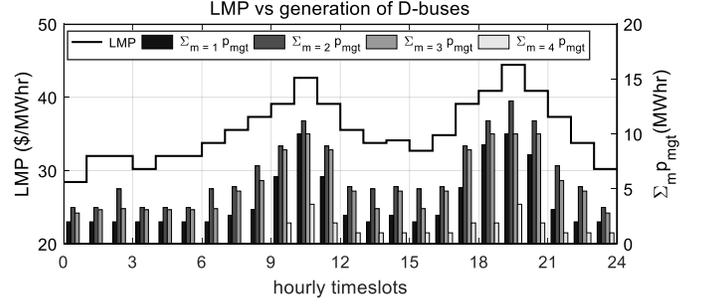

Fig. 3. Hourly D-bus generation vs. LMP

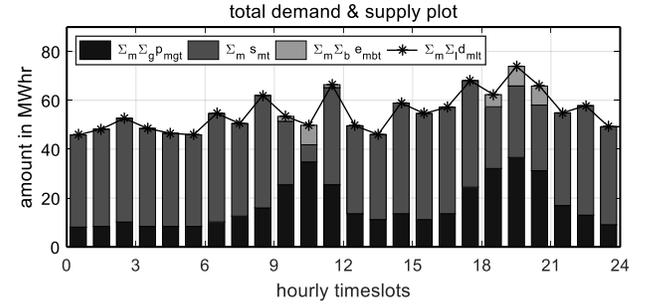

Fig. 4. Total demand and supply by the DSO, BSSs and generators

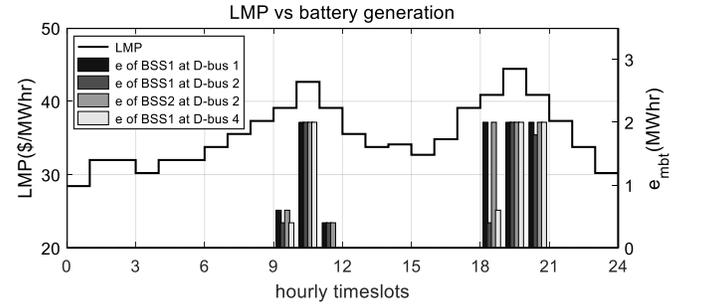

Fig. 5. Supply from BSSs as DLMP changes

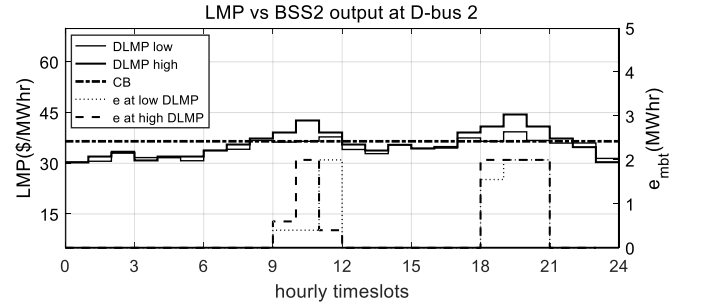

Fig. 6. Supply from BSS units at two different DLMPs

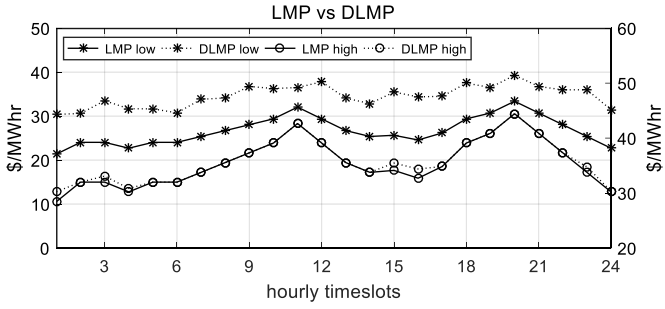

Fig. 7. LMP versus DLMP for 24-hour horizon

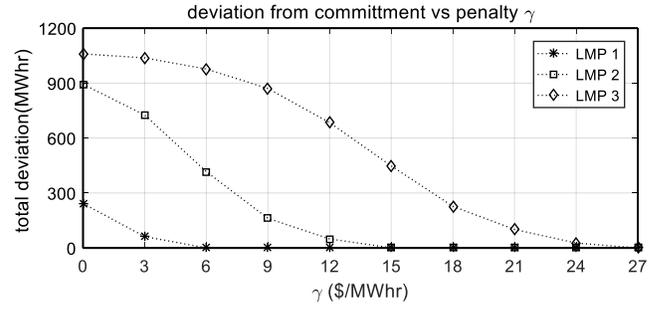

Fig. 9. Deviation from ISO supply vs penalty for three different LMPs, with $LMP1(t) < LMP2(t) < LMP3(t)$

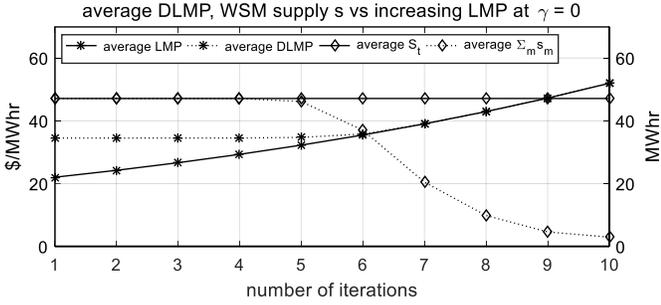

Fig. 8. Average LMP vs DLMP and total supply to DSO

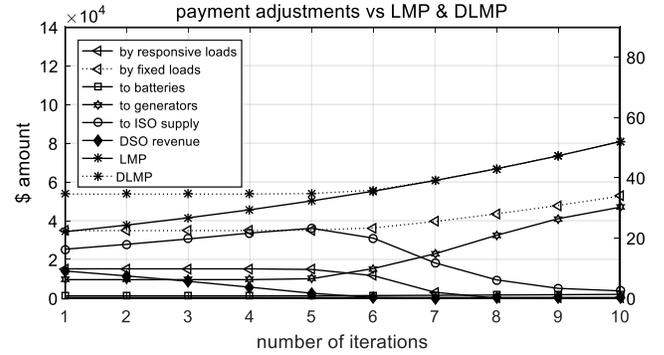

Fig. 10. Payments and reimbursements

Fig. 5 depicts the LMP values versus BSS units' commitment considering its charging/discharging limits. All BSS units are set to 6 hours of minimum charging and three minimum discharging hours except the minimum discharge time of BSS at D-bus 4, which is set to 2 hours. Note that this plot meets all the requirements set in the constraints in Eqns. (20) – (30). BSS units are scheduled during high price hours with more power scheduled at peak LMPs than its neighboring hours. In addition, the sum of assigned energy during scheduling horizon does not exceed each BSS unit's capacity. The plot in Fig.6 shows scheduled behavior of the second BSS unit at D-bus 2 for two different DLMPs along with its declared selling price depicted with the horizontal line. As seen, the BSS is only scheduled when DLMP is higher than its bid. Furthermore, the scheduled amount of power withdrawal from this BSS is higher where the difference between its bid and the DLMP is higher. A similar observation was made when generation units' schedules were investigated.

In order to study the dynamics of DLMP versus LMP, Figs. 7 and 8 were plotted. Fig.7 shows the change in DLMP during the scheduling horizon for two different sets of LMPs. When LMP is low compared to the bids of loads, generation, and BSS units, DLMP deviates significantly and becomes higher than LMP. In the case when LMP increases, the DLMP also increases and overlaps with LMP in most hours. Note that further increase in LMP causes DLMP to equal LMP in all hours. This is because at LMPs higher than the generation units' bids, all generation units are scheduled to serve the fixed and or price responsive loads, and serving any extra MWhr incurs a cost equal to LMP. Fig.8 illustrates this concept further by showing the 24-h LMP average for 10 scenarios while increasing LMP by a fixed percentage at each scenario. As expected, at low LMPs the entire committed supply of the WSM is injected into the D-buses, whereas lower amounts are drawn when LMP increases.

The simulation results reported here assumes that no penalty is incurred for any deviation from what is committed to ISO. The DSO's objective function in Eqn. (1) can be modified by adding a linear penalty function $\phi(\gamma)$ to account for penalty incurred due to any deviation from original commitment.

$$\phi(\gamma) = -\gamma \left( \sum_t S_t - \sum_t \sum_m s_{mt} \right) \quad (31)$$

To show the effect of applying penalty, the total deviation over the scheduling horizon from original commitment was plotted as a function of $\gamma$ in Fig.9 for three different LMPs. Notice that for higher LMPs, a higher penalty is required to make the deviation zero. This means that if LMP is high, more internal generation and BSS units are scheduled, and it takes a higher penalty to force power injection from the WSM in order to make the deviation zero.

Fig. 10 depicts the budget dynamics of the 10 increasing LMP scenarios for a fixed commitment $S_t$ from WSM when no penalty for deviation is applied, i.e. $\gamma = 0$. At lower LMPs, during scenarios one to five, the DSO makes money. It sells energy at higher DLMP while buying it at lower LMP. As the LMP is increased further, less power is purchased from the WSM and DLMP approaches LMP. As a result, the DSO's revenue drops down to zero after the fifth scenario. Note that, if DSO is penalized for deviation, it loses a monetary amount equal to $\phi(\gamma)$ given by Eqn. (31). This is because deviation occurs at higher LMPs, i.e. after scenario five, when DSO's revenue is zero with no penalty ($\gamma = 0$). In such scenarios, the DSO has to bear the deviation cost $\phi(\gamma)$.

## IV. CONCLUSION

In this paper, the distribution system operator's DAM auction in the presence of distribution level generation units, renewable energy resources coupled with BSS units, and loads with fixed and price responsive segments has been modeled and studied. The DSO is considered to have the knowledge of its supply amount at LMP from the WSM. The DSO uses MILP to optimally schedule its available resources and maximize the social welfare. By clearing the auction at DLMP, the dynamics of DLMP versus LMP and their effect on the outcome of the auction as well as the resulting payments were studied. Simulation results show that, if DSO is not penalized for deviating from its committed schedule with the WSM, the auction is always weakly budget balanced. The DSO only makes money when LMP is cheaper at a given fixed supply from the WSM.

Future studies can be carried out to model the auction in an iterative and distributed manner [16], [17]. The authors are working on establishing DLMPs locally using a lower level auction at each D-bus by achieving general market equilibrium conditions. Inclusion of distribution system physical constraints such as bus voltages, lines flow, and transformers capacity constraints in a distributed manner within a decentralized version of the proposed auction framework can be another interesting line of research.

## REFERENCES

[1] Verzijlbergh, R. A., Z. Lukszo, and M. D. Ilić. "Comparing different EV charging strategies in liberalized power systems." In 9*th International Conference on the European Energy Market*, pp. 1-8, 2012.

[2] Verzijlbergh, Remco A., Laurens J. De Vries, and Zofia Lukszo. "Renewable energy sources and responsive demand. Do we need congestion management in the distribution grid?" IEEE Transactions on Power Systems, vol. 29, no. 5, pp. 2119-2128, 2014.

[3] Fitzgerald, Garrett, James Mandel, Jesse Morris, and Hervé Touati. "The Economics of Battery Energy Storage", *Rocky Mountain Institute*, September 2015.

[4] Mitra, Joydeep, and Mallikarjuna R. Vallem. "Determination of storage required to meet reliability guarantees on island-capable microgrids with intermittent sources." *IEEE Transactions on Power Systems*, vol. 27, no. 4, pp.2360-2367, 2012.

[5] Rahimi, F., and S. Mokhtari. "From ISO to DSO: imagining new construct--an independent system operator for the distribution network." *Public Util. Fortn,* vol.152, no. 6, pp. 42-50, 2014.

[6] Parhizi, Sina, and Amin Khodaei. "Market-based microgrid optimal scheduling." *IEEE International Conference on Smart Grid Communications (SmartGridComm)*, pp. 55-60, 2015.

[7] Kristov, L., and P. De Martini. "21st century electric distribution system operations." pp. 1-11, May 2014.

[8] Khodaei, Amin. "Microgrid optimal scheduling with multi-period islanding constraints." *IEEE Transactions on Power Systems*, vol.29, no. 3, pp. 1383-1392, 2014.

[9] Gkatzikis, Lazaros, Iordanis Koutsopoulos, and Theodoros Salonidis. "The role of aggregators in smart grid demand response markets." *IEEE Journal on Selected Areas in Communications*, vol.31, no. 7, pp.1247-1257, 2013.

[10] Majumder, Bodhisattwa P., M. Nazif Faqiry, Sanjoy Das, and Anil Pahwa. "An efficient iterative double auction for energy trading in microgrids." *IEEE Symposium on Computational Intelligence Applications in Smart Grid (CIASG)*, pp. 1-7, 2014.

[11] Sbordone, Danilo Antonio, Enrico Maria Carlini, Biagio Di Pietra, and Michael Devetsikiotis. "The future interaction between virtual aggregator-TSO-DSO to increase DG penetration." *International Conference on Smart Grid and Clean Energy Technologies (ICSGCE)*, pp. 201-205, 2015.

[12] Faqiry, M. Nazif, and Sanjoy Das. "Double-Sided Energy Auction in Microgrid: Equilibrium under Price Anticipation." IEEE Access, vol.4, pp.3794-3805, 2016.

[13] Li, Ruoyang, Qiuwei Wu, and Shmuel S. Oren. "Distribution locational marginal pricing for optimal electric vehicle charging management." *IEEE Transactions on Power Systems* vol. 29, no. 1, pp. 203-211, 2014.

[14] Huang, Shaojun, Qiuwei Wu, Shmuel S. Oren, Ruoyang Li, and Zhaoxi Liu. "Distribution locational marginal pricing through quadratic programming for congestion management in distribution networks." *IEEE Transactions on Power Systems*, vol.30, no. 4, pp. 2170-2178, 2015.

[15] O'Connell, Niamh, Qiuwei Wu, Jacob Østergaard, Arne Hejde Nielsen, Seung Tae Cha, and Yi Ding. "Day-ahead tariffs for the alleviation of distribution grid congestion from electric vehicles." *Electric Power Systems Research,* vol.92, pp. 106-114, 2012.

[16] Faqiry, M. Nazif, and Sanjoy Das. "Transactive Energy Auction with Hidden User Information in Microgrid." arXiv preprint arXiv: 1608.03649, 2016.

[17] Miyamoto, Toshiyuki, Kazuyuki Mori, Shoichi Kitamura, and Yoshio Izui. "Solving Distributed Unit Commitment Problem With Walrasian Auction." *IEEE Transactions on Systems, Man, and Cybernetics: Systems* vol. 46, no. 8, 2016.

[18] Sotkiewicz, Paul M., and Jesus M. Vignolo. "Nodal pricing for distribution networks: efficient pricing for efficiency enhancing DG." *IEEE transactions on Power Systems*, vol. 21, no. 2, pp. 1013, 2006.

[19] Baran, M. E., and Felix F. Wu. "Optimal sizing of capacitors placed on a radial distribution system." *IEEE Transactions on power Delivery*, vol.4, no.1, pp. 735-743, 1989.

[20] Wang, Zhaoyu, Bokan Chen, and Jianhui Wang. "Decentralized energy management system for networked microgrids in grid-connected and islanded modes." *IEEE Transactions on Smart Grid*, vol.7, no.2, pp.1097-1105, 2016.